\documentclass[runningheads]{llncs}
\usepackage{graphicx}
\usepackage{authblk}
\usepackage{graphicx}
\usepackage{subfigure}
\usepackage{array}
\usepackage{float}
\usepackage{booktabs}
\usepackage{bbding}
\usepackage{etoolbox}
\usepackage{color}
\usepackage{soul}
\usepackage{algorithm}
\usepackage{algorithmic}
\usepackage{cite}
\usepackage{balance}
\usepackage{amsmath, amssymb}
\usepackage{makecell}
\usepackage{multirow}
\usepackage{hyperref}

\begin{document}
\title{EndoUIC: Promptable Diffusion Transformer for Unified Illumination Correction in Capsule Endoscopy}
\titlerunning{EndoUIC: Unified Illumination Correction in Capsule Endoscopy}

\authorrunning{L. Bai et al.}

\author{Long Bai\inst{1,2~\star}
\and Tong Chen\inst{3~\star}
\and Qiaozhi Tan\inst{2}
\thanks{Co-first authors.}
\and Wan Jun Nah\inst{2}
\and Yanheng Li\inst{4}
\and Zhicheng He\inst{2}
\and Sishen Yuan\inst{1}
\and Zhen Chen\inst{5}
\and Jinlin Wu\inst{5}
\and Mobarakol Islam\inst{6}
\and Zhen Li\inst{7}
\and Hongbin Liu\inst{5}
\and Hongliang Ren\inst{1,2}
\thanks{Corresponding author.}}

\institute{Department of Electronic Engineering, The Chinese University of Hong Kong (CUHK), Hong Kong, China
\and Shenzhen Research Institute, CUHK, Shenzhen, China
\and The University of Sydney, Sydney, NSW, Australia
\and City University of Hong Kong, Hong Kong, China
\and Centre for Artificial Intelligence and Robotics, HKISI-CAS, Hong Kong, China
\and University College London, London, UK
\and Qilu Hospital of Shandong University, Jinan, China\\
\email{b.long@link.cuhk.edu.hk, hlren@ee.cuhk.edu.hk}
}

\maketitle              
\begin{abstract}
Wireless Capsule Endoscopy (WCE) is highly valued for its non-invasive and painless approach, though its effectiveness is compromised by uneven illumination from hardware constraints and complex internal dynamics, leading to overexposed or underexposed images. While researchers have discussed the challenges of low-light enhancement in WCE, the issue of correcting for different exposure levels remains underexplored. To tackle this, we introduce EndoUIC, a WCE unified illumination correction solution using an end-to-end promptable diffusion transformer (DiT) model. In our work, the illumination prompt module shall navigate the model to adapt to different exposure levels and perform targeted image enhancement, in which the Adaptive Prompt Integration (API) and Global Prompt Scanner (GPS) modules shall further boost the concurrent representation learning between the prompt parameters and features. Besides, the U-shaped restoration DiT model shall capture the long-range dependencies and contextual information for unified illumination restoration. Moreover, we present a novel Capsule-endoscopy Exposure Correction (CEC) dataset, including ground-truth and corrupted image pairs annotated by expert photographers. Extensive experiments against a variety of state-of-the-art (SOTA) methods on four datasets showcase the effectiveness of our proposed method and components in WCE illumination restoration, and the additional downstream experiments further demonstrate its utility for clinical diagnosis and surgical assistance. The code and the proposed dataset are available at \href{https://github.com/longbai1006/EndoUIC}{github.com/longbai1006/EndoUIC}.

\end{abstract}
\section{Introduction}

Wireless Capsule Endoscopy (WCE) has revolutionized gastrointestinal (GI) diagnostics by offering a minimally invasive, painless way of examination in the GI tract~\cite{zhang2022deep}. 
However, the effectiveness of WCE can often be influenced due to factors such as limited battery capacity, camera performance, and the complexity of the GI tract~\cite{wang2023rethinking}. Uneven illumination within the tract can significantly degrade image quality, thus affecting the accuracy and efficiency of diagnosis, screening, and the provision of timely feedback~\cite{yue2023deep}. 
While the issue of low-light image enhancement (LLIE) in WCE images has received considerable attention, leading to various strategies to improve visibility in low-light areas~\cite{long2018adaptive,moghtaderi2024endoscopic}, the challenge of overexposure remains less explored~\cite{yang2023learning}. Various solutions~\cite{bai2023llcaps, mou2023global,ma2020cycle,yue2023deep} have been put forward to enhance low-light WCE images. Nevertheless, the complex and dynamic internal body environment will also result in overexposure, which obscures critical details with excessive brightness, as the brightness levels often extend beyond the dynamic range these techniques can adequately adjust~\cite{rukundo2017advanced}. 

Conventional approaches have been utilized to enhance the structure visibility in WCE images~\cite{rukundo2017advanced,wang2022endoscopic}. However, compared to deep learning methods, they tend to be less adaptive, less content-aware, and require manual intervention. Sequentially, García-Vega \emph{et al.}~\cite{garcia2023multi} implemented a structure-aware deep neural network for exposure correction (EC), and employed CycleGAN~\cite{CycleGAN2017} for EC dataset generation~\cite{garcia2022novel}. Presently, solutions for WCE unified illumination adaptation are still underexplored, lacking an end-to-end architecture that can unify illumination correction tasks. Furthermore, existing endoscopic EC datasets produced via generative models struggle to replicate the complexity encountered in real-world scenarios. This gap underscores the need for a unified light adaptation model capable of concurrently tackling over \& underexposed images, which is crucial for the retention and enhancement of vital diagnostic details.

Denosing diffusion probabilistic models (DDPMs) have demonstrated outstanding performance in low-level vision tasks including denoising, super resolution, and low-light enhancement, owing to their ability to model complex data distributions and incorporate conditional information effectively~\cite{ho2020DDPM,wang2023learning}. In scenarios involving over \& underexposed images, which typically demand different parameter spaces and optimization trajectories, directly training diffusion models might not be the best approach. Contrastive learning methods have already been employed to learn varying image degradation types, while an additional network would be needed~\cite{li2022air}. To this end, we introduce a set of learnable parameters that act as our prompt. These prompt parameters are optimized through an end-to-end process, learning to adjust the model's prior for different image degradation. Then, it shall steer the model within the parameter space toward different low-level details essential for EC and LLIE. Thus, leveraging the task-specific knowledge acquired by the model, it dynamically adapts the input data according to different brightness levels. Additionally, the prompt module can act as an attention mechanism and increase the depth of the network. Therefore, even if the input exhibits only one type of degradation, the model can still maintain effective restoration performance.
Moreover, to address the issue of data scarcity, we have collected a WCE dataset and invited photography experts to annotate over \& underexposed images manually. Specifically, our contributions to this work can be summarized as three-fold: 

\begin{itemize}
    \item [--]We propose \textbf{EndoUIC} - \textbf{Endo}scopic \textbf{U}nified \textbf{I}llumination \textbf{C}orrection - a promptable diffusion model for unified WCE illumination correction. Specifically, the illumination prompt module is designed to navigate the diffusion model toward specific illumination conditions. 
    \item [--]In our proposed framework, we embed a diffusion process within a U-shape transformer to perceive global illumination and multiscale contextual information, and utilize prompts to guide the illumination restoration procedure. Our prompt module contains an Adaptive Prompt Integration (API) module, which dynamically produces and integrates prompt parameters with feature representations. Additionally, we incorporate the Global Prompt Scanner (GPS)  module to enhance the interaction between prompts and features.
    \item [--]To tackle the data shortage issue, we propose a novel WCE EC dataset, named \textbf{C}apsule-endoscopy \textbf{E}xposure \textbf{C}orrection (\textbf{CEC}) dataset, with normal and wrongly exposed image pairs. Extensive comparison, ablation, and downstream experiments on four datasets demonstrate the superior effectiveness of our EndoUIC, showcasing its potential in clinical applications.
\end{itemize}

\section{Methodology}

\subsection{Preliminaries} 

\subsubsection{Visual Prompt Learning} introduces a set of learnable parameters that provide deep learning models with contextual information regarding the image degradation types in image restoration tasks~\cite{li2023prompt,potlapalli2023promptir}. These prompts interact with the features of the input image, directing the model to adaptively adjust to different degradation types, thus restoring high-quality and clean images. This method enables a single unified model to address multiple image degradation challenges, enhancing the model's generalization capabilities.

\subsubsection{Pyramid Diffusion Models} (PyDiff) is an LLIE diffusion model that implements a pyramid diffusion strategy~\cite{zhou2023pyramid}. Unlike DDPMs, where image resolution remains constant throughout the reverse process, PyDiff starts with a lower resolution and progressively increases it to a higher resolution in the diffusion process. The forward and reverse process can be formulated with the given input $\mathbf{x}_0$, time step $t$, noise schedule $\{\alpha\}_{t=0}^T$, and scaling schedule $\{U\}_{t=0}^T$:
\begin{equation}
\resizebox{.55\linewidth}{!}{$
    q\left(\mathbf{x}_t \mid \mathbf{x}_{t-1}\right)=\mathcal{N}\left(\mathbf{x}_t ; \sqrt{\bar{\alpha}_t}\left(\mathbf{x}_0 \downarrow_{U_t / U_{t-1}}\right),\left(1-\bar{\alpha}_t\right) \mathbf{I}\right)
$}
\end{equation}
\begin{equation}
\resizebox{\linewidth}{!}{$
    p_\theta\left(\mathbf{x}_{t-1} \mid \mathbf{x}_t\right)= \begin{cases}\mathcal{N}\left(\mathbf{x}_{t-1} ; \frac{\sqrt{\alpha_{t-1}} (1-\alpha_t)}{1-\bar{\alpha}_t} \mathbf{y}_\theta\left(\mathbf{x}_t\right)+\frac{\sqrt{\alpha_t}\left(1-\bar{\alpha}_{t-1}\right)}{1-\bar{\alpha}_t} \mathbf{x}_t\right. \left. , \frac{1-\bar{\alpha}_{t-1}}{1-\bar{\alpha}_t} (1-\alpha_t) \mathbf{I}\right), & \text { if } U_t=U_{t-1} \\ \mathcal{N}\left(\mathbf{x}_{t-1} ; \sqrt{\bar{\alpha}_{t-1}}\left(\mathbf{y}_\theta\left(\mathbf{x}_t\right) \uparrow_{U_t / U_{t-1}}\right)\right. \left.,\left(1-\bar{\alpha}_{t-1}\right) \mathbf{I}\right), & \text { if } U_t>U_{t-1}\end{cases}
$}
\end{equation}
in which $\alpha_t \in(0,1)$ and $\bar{\alpha}_t= \prod_{i=1}^t \alpha_i$. While $a_t \geq a_{t+1}$ is getting bigger noise, $s_t \leq s_{t+1}$ is getting lower resolution.
This approach optimizes the sampling speed and achieves improved image restoration quality by gradually refining image details with increasing resolution. 

\subsection{Proposed Method: EndoUIC}

\begin{figure*}[t]
    \centering
    \includegraphics[width=1\linewidth, trim=5 140 50 0]{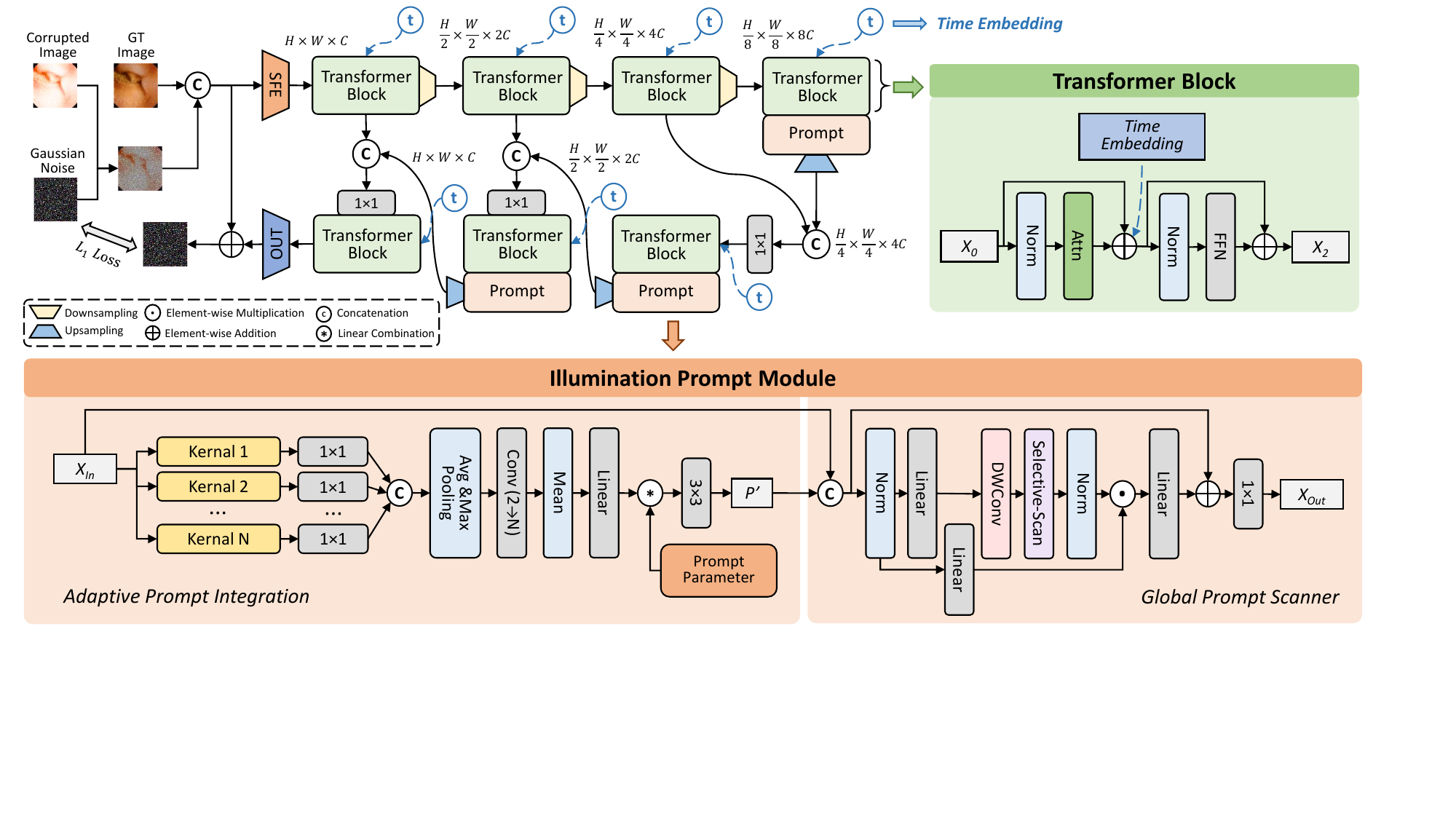}
    \caption{The overview of our EndoUIC. Our network comprises the 4-level diffusion transformer (DiT), which is used to predict the noise. In each upsampling stage of the restoration DiT, the illumination prompt module is incorporated, which consists of the Adaptive Prompt Integration (API) and a Global Prompt Scanner (GPS) blocks. 
    `SFE' and `OUT' denote the shallow feature extractor and the output block, respectively.
    }
    \label{fig:mainframe}
\end{figure*}
Our proposed EndoUIC framework is presented in Fig.~\ref{fig:mainframe}, with the U-shape restoration DiT to estimate noises. The network is optimized with simple $\mathcal{L}_1$ loss and the noise sampling strategy follows~\cite{zhou2023pyramid}.
\\
\\
\noindent\textbf{Restoration Diffusion Transformer.}
Our network begins with a shallow feature extractor that transforms the image into the feature representation $R$. The features are then fed into the 4-level down-sampling transformer encoder and up-sampling transformer decoder, which is similar to the UNet structure~\cite{ronneberger2015u}. The skip connections are executed at each level of the encoder-decoder. With $\mathcal{X}_0$ and $\mathcal{X}_{2}$ denoting the input and output feature respectively, each transformer block can be formulated as:
\begin{equation}
    \mathcal{X}_1 = Attn({Norm}(\mathcal{X}_0)), \;\; \mathcal{X}_{2} = FFN({Norm}(\mathcal{X}_1))
\end{equation}
in which the time embedding $t$ is injected with $\mathcal{X}_1$. $FFN$ denotes the Feed-Forward Network. The decoder finally outputs a high-resolution image with normal illumination after propagating through the output block.
Each up-sampling stage incorporates an illumination prompt module. Firstly, the output of the prompt module is concatenated with the corresponding encoder's skip connection output. Then it will be passed through a $1 \times 1$ convolutional (Conv) layer before being input into the respective decoder.

\noindent\textbf{Illumination Prompt Module.}
Our illumination prompt module shall utilize additional prompt parameters to encode key discriminative information about brightness degradation levels. The Adaptive Prompt Integration (API) block is used to learn discriminative illumination information, and the Global Prompt Scanner (GPS) integrates the learned prompt parameters with the image features, guiding the model's learning process. Given $P$ as the learnable prompt parameters, $\mathcal{X}_{In}$ as the input features, and $\mathcal{X}_{Out}$ as the output features, the prompt module can be formulated:
\begin{equation}
    P' = \mathcal{F}_{API}(\mathcal{X}_{In};P), \;\; \mathcal{X}_{Out} = \mathcal{F}_{GPS}(\mathcal{X}_{In}; P')
\end{equation}

\noindent\textbf{Adaptive Prompt Integration.}
The API module is designed to generate the prompt parameters and integrate them with the adaptively learned feature maps. We first define a set of learnable parameters $P$, which are designed to embed different illumination conditions into the features. This design can efficiently capture long-range dependencies to perceive global illumination information and address local-region uneven illumination conditions. Thus, our method can also effectively learn the illumination representation of features.

To achieve this, we refrain from directly multiplying $P$ with the features as this could diminish the correlation between $P$ and the features. Instead, we employ a multi-scale dynamic feature space for efficient feature learning. Specifically, we construct kernels of varying sizes to equip the model with different receptive fields~\cite{li2023large}, and fuse the feature representations using the $1\times1$ layer.

As illustrated in the left-bottom of Fig.~\ref{fig:mainframe}, our dynamic kernel selection mechanism concatenates features from different receptive fields to obtain the combined feature $\mathcal{X_A}$. We then utilize average and max pooling to extract spatial information from the feature space. The extracted features, after concatenation, are passed through a Conv layer, which expands the channel from $2$ to $\mathbf{N}$:
\begin{equation}
\mathcal{X}_A' = \mathcal{F}_{2 \rightarrow \mathbf{N}}[AvgPool(\mathcal{X_A}) \| MaxPool(\mathcal{X_A})]
\end{equation}
where $[\cdot\|\cdot]$ denotes concatenation. Subsequently, we apply the Sigmoid activation $\sigma$ to obtain the weighted coefficient, which is then multiplied with $P$ using the following equation, enabling dynamic feature learning to weight $P$ adaptively:
\begin{equation}
P' = \mathcal{F}_{FCN}[Mean(\sigma(\mathcal{X_A}'))] \odot P
\end{equation}
where $\odot$ denotes element-wise multiplication and $\mathcal{F}_{FCN}$ means the fully-connected layer. The obtained $Conv_{3\times3}(P')$ is then propagated through the GPS module for further correlation learning of the feature maps and prompt parameters.

\noindent\textbf{Global Prompt Scanner.}
In the GPS module, the prompt parameter $P'$ is firstly concatenated with the input feature $\mathcal{X}_{In}$, utilizing $P'$ to guide the process of luminance restoration:
\begin{equation}
    \mathcal{X}_P = [\mathcal{X}_{In}\|P']
\end{equation}

The selective-scan mechanism of VMamba~\cite{liu2024vmamba}, which captures long-range representations by scanning sequentially from four directions (top-left $\rightarrow$ bottom-right, bottom-right $\rightarrow$ top-left, top-right $\rightarrow$ bottom-left, bottom-left $\rightarrow$ top-right), has proven to be an effective approach for learning visual representations. 
To further enhance the global perception and foster the interaction between $\mathcal{X}_{In}$ and $P'$, we conduct the cross-scan on $\mathcal{X}_P$. In this case, the scans in the same dimension as the concatenation can effectively facilitate the interaction between $P'$ and $\mathcal{X}_{In}$, while the scans vertical to the concatenation dimension can promote the internal representation learning within $P'$ and $\mathcal{X}_{In}$ themselves.

The GPS module is presented in the right-bottom of Fig.~\ref{fig:mainframe}(c). Features that follow a skip connection are processed by a $1 \times 1$ Conv layer and a $3 \times 3$ Conv layer. The combined feature is then fused with features from higher spatial dimensions to facilitate the illumination restoration process of the overall model.

\section{Experiments}
\label{sec:exper}

\begin{figure*}[t]
    \centering
    \includegraphics[width=0.7\linewidth, trim=0 1070 520 50]{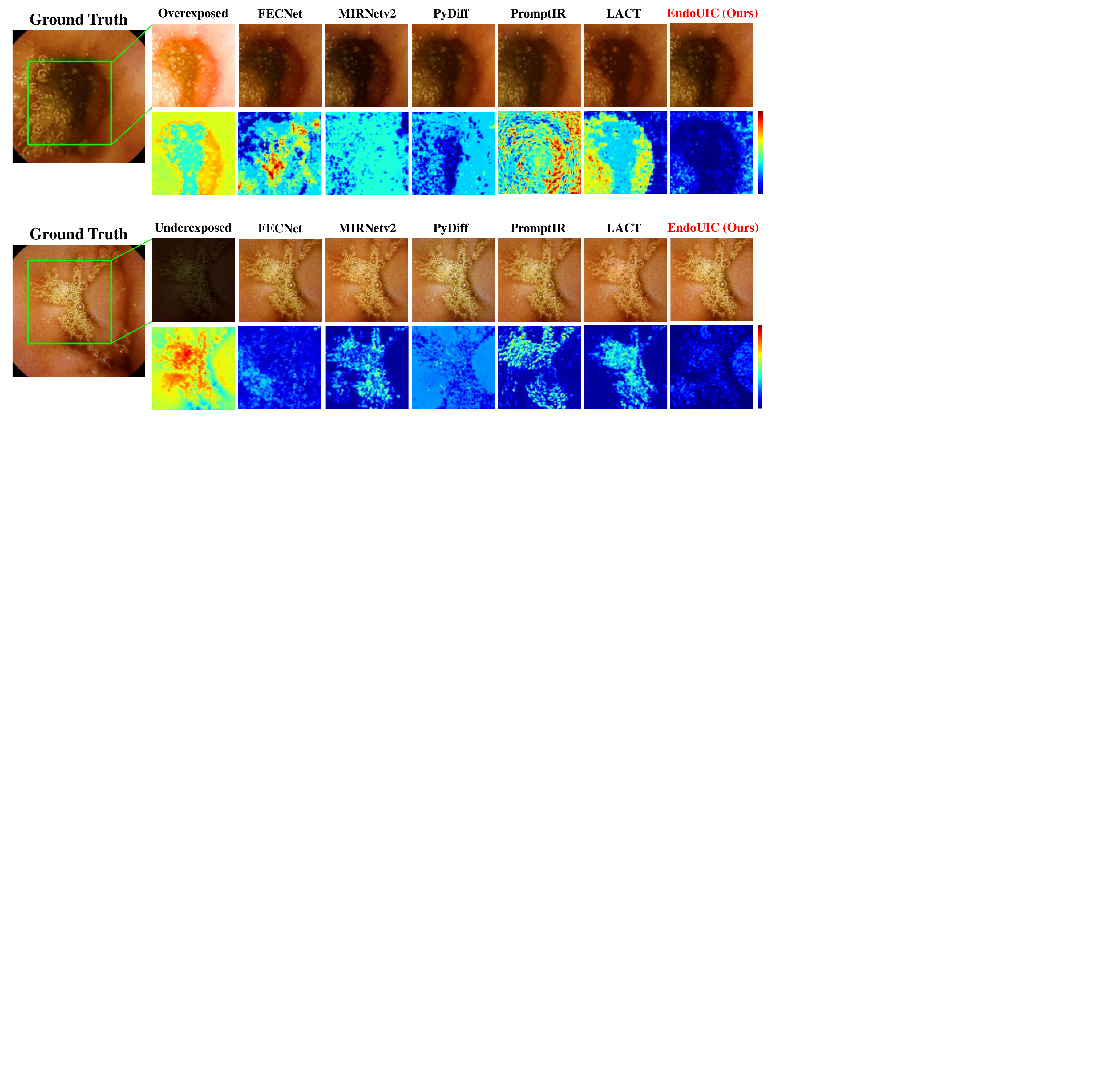}
    \caption{The visualization results and error maps of our EndoUIC against SOTA methods on the CEC dataset. We present the enhanced images by SOTA exposure correction methods and their heat maps of reconstruction errors, with blue indicating lower errors and red denoting higher errors.
    }
    \label{fig:IMG_visual}
\end{figure*}

\subsection{Dataset}
\label{sec:dataset}

We conduct our experiments on two EC datasets and two LLIE datasets:

\noindent \textbf{Capsule endoscopy Exposure Correction (CEC)} dataset is collected by ANKON magnetically controlled WCEs of three patients. The training set includes 800 images from two patients, and the test set contains 200 images from another patient. The dataset comprises half overexposed and half underexposed images. 
We invite expert photographers to simulate the camera aperture settings with the Adobe Camera Raw SDK by adjusting exposure values (EVs). Changing the exposure value is equivalent to changing the camera aperture size. We render the raw image with varying numerical EVs to change the exposure range of highlights and shadows, emulating real exposure mistakes.

\noindent \textbf{Endo4IE} dataset is a public synthetic EC dataset of conventional endoscopy~\cite{garcia2022novel}. It was created by initially selecting public images without exposure issues. Then, CycleGAN~\cite{CycleGAN2017} was applied to generate paired over \& underexposed synthetic images, and MSE and SSIM metrics were used to filter and finalize a dataset of $985$ underexposed and $1231$ overexposed images.

\noindent \textbf{Kvasir-Capsule}~\cite{smedsrud2021kvasir} and \textbf{Red Lesion Endoscopy (RLE)}~\cite{coelho2018RLE} are originally two datasets utilized for WCE disease diagnosis. Bai \emph{et al.}~\cite{bai2023llcaps} have curated images from these datasets and synthesized two datasets specifically tailored for WCE LLIE by applying random Gamma correction and illumination reduction. Specifically, the Kvasir-Capsule dataset comprises 2000 training images and 400 test images. The RLE dataset contains 946 training images and 337 test images.

\subsection{Implementation Details}
\label{sec:implementation}

The performance of our proposed EndoUIC is compared with a variety of state-of-the-art (SOTA) LLIE and EC methodologies, which are listed in Table~\ref{tab:1}, \ref{tab:2}, \ref{tab:3}, and supplementary materials. For methods marked with `*', we obtain results directly from previous works. For the remaining methods, we reproduce the results through their official repositories.
We conduct our experiments with Python PyTorch on NVIDIA A100 GPUs.
We train our model with Adam for $1000$ epochs. The learning rate 
is set to $10^{-4}$. 
We evaluate the image enhancement performance with Peak Signal-to-Noise Ratio (PSNR), Structural Similarity Index (SSIM), and Learned Perceptual Image Patch Similarity (LPIPS). We also follow the previous work~\cite{bai2023llcaps} to conduct a downstream medical diagnosis task on the RLE test set - the red lesion segmentation. The UNet~\cite{ronneberger2015u} is trained using Adam with 20 epochs and a learning rate of $10^{-4}$, and evaluated with mIoU.

\begin{table}[t]
	\caption{
        EC comparison against existing and SOTA methods on our CEC dataset.
	}
 	\centering
	\label{tab:1}  
 \resizebox{\textwidth}{!}{	
\begin{tabular}{c|ccccccccccc}
\hline
Methods & FECNet~\cite{huang2022deep} & SID~\cite{huang2022exposure} & DRBN~\cite{huang2022exposure} & MIRv2~\cite{zamir2022MIRNetv2} & LLCaps~\cite{bai2023llcaps} & PyDiff~\cite{zhou2023pyramid} & PromptIR~\cite{potlapalli2023promptir} & LACT~\cite{baek2023luminance} & PIP~\cite{li2023prompt} & EndoUIC  \\ \hline
PSNR $\uparrow$ & 28.78 & 24.29 & 26.83 & 28.36 & 27.55 & 28.18 & 28.27 & 28.40 & 25.01 & \textbf{29.65} \\ 
SSIM $\uparrow$ & 92.61 & 85.69 & 90.50 & 93.58 & 85.95 & 95.79 & 83.14 & 93.09 & 70.09 & \textbf{96.80} \\ 
LPIPS $\downarrow$ & 0.1048 & 0.2111 & 0.1452 & 0.1080 & 0.2366 & 0.0941 & 0.0717 & 0.1103 & 0.1527 & \textbf{0.0655}\\ \hline
\end{tabular}}
\end{table}

\begin{table}[t]
	\caption{
        EC comparison against existing and SOTA methods on the Endo4IE dataset. `*' means we use the results from the previous works.
        }
 	\centering
	\label{tab:2}  
 \resizebox{\textwidth}{!}{	
\begin{tabular}{c|ccccccccccc}
\hline
\multicolumn{1}{c|}{Methods} & LMSPEC*~\cite{garcia2022novel} & LMSPEC+*~\cite{garcia2023multi} & FECNet~\cite{huang2022deep} & MIRv2~\cite{zamir2022MIRNetv2} & LA-Net~\cite{yang2023learning} & PyDiff~\cite{zhou2023pyramid} & PromptIR~\cite{potlapalli2023promptir} & LACT~\cite{baek2023luminance} & PIP~\cite{li2023prompt} & EndoUIC  \\ \hline
PSNR $\uparrow$ & 23.97 &23.62  &24.72  &23.85  & 23.51  &24.73  & 23.73  & 22.92 & 25.28  &\textbf{25.49}  \\ 
SSIM $\uparrow$ & 80.34 &79.97  &81.84  &82.33  & 83.78  &84.78  & 79.57 & 76.88  & 81.94  &\textbf{85.20}  \\ 
LPIPS $\downarrow$ & -  & -  &0.2031  &0.2376  & \textbf{0.1186} & 0.2148  & 0.2396  & 0.2671 & 0.2150  &0.1937 \\ \hline
\end{tabular}}
\end{table}

\subsection{Results}
\label{sec:results}
\noindent\textbf{EC Comparison.} As indicated in Tables~\ref{tab:1} and \ref{tab:2}, we initially perform the endoscopy exposure correction experiment in comparison with the existing SOTA methods. Our approach successfully surpasses various methods with different architectures (e.g., CNNs, Transformers, and DDPMs). %
Our method primarily restores images at the pixel level, leading to top-1 results in PSNR and SSIM. However, it falls behind LANet in feature perception quality, while our method still ranks 2nd in LPIPS. Overall, our EndoUIC achieves the best results on our self-proposed CEC dataset and the public Endo4IE benchmark. 
The visualization of our image enhancement results and their corresponding error maps are presented in Fig.~\ref{fig:IMG_visual}, where blue indicates fewer errors. It is observable that our method demonstrates the least errors and optimal results.
Additionally, our EndoUIC demonstrates the FLOPs number of 14.36M and a standard deviation of 3.51 in PSNR on the CEC dataset. Future work will focus on enhancing the inference speed and robustness of our proposed framework.

\begin{table}[t]
	\caption{
        LLIE comparison with existing and SOTA solutions on the Kvasir-Capsule~\cite{smedsrud2021kvasir} and RLE datasets~\cite{coelho2018RLE}. The red lesion segmentation experiment is conducted on the RLE test set~\cite{coelho2018RLE} by following the previous work~\cite{bai2023llcaps}. LLCaps* means we use the results from the previous SOTA~\cite{bai2023llcaps}.
	}
 	\centering
	\label{tab:3}  
\resizebox{1\textwidth}{!}{	
\begin{tabular}{cc|ccccccccc}
\hline
\multicolumn{2}{c|}{Methods} & LLCaps*~\cite{bai2023llcaps}& PIP~\cite{li2023prompt}&CFWD~\cite{xue2024low}&Diff-LOL~\cite{jiang2023low}&LA-Net~\cite{yang2023learning}& CLE~\cite{yin2023cle}&PyDiff~\cite{zhou2023pyramid}&PromptIR~\cite{potlapalli2023promptir}&EndoUIC\\\hline
\multirow{3}{*}{\makecell[c]{Kvasir- \\ Capsule}}  & \multicolumn{1}{|c|}{PSNR $\uparrow$} &35.24 &33.60&35.88&33.60&30.84&26.55&35.07&33.54& \textbf{36.85}        \\\cline{2-11}
&\multicolumn{1}{|c|}{SSIM $\uparrow$} &96.34 &95.09&96.26&95.42&95.32&87.87&96.60&96.77&\textbf{97.04}         \\\cline{2-11}
& \multicolumn{1}{|c|}{LPIPS $\downarrow$} &0.0374 &0.0302&0.0467&0.0847&0.0562&0.0829&0.0364&0.0377&\textbf{0.0255}         \\\hline

\multirow{3}{*}{RLE}  & \multicolumn{1}{|c|}{PSNR $\uparrow$} & 33.18 & 28.60 & 30.14 & 28.46 & 25.92 & 26.20 & 33.21 & 32.07 & \textbf{33.50}           \\\cline{2-11}
&\multicolumn{1}{|c|}{SSIM $\uparrow$} & 93.34 & 87.27 & 90.25 & 82.52 & 85.72 & 81.42 & 93.54&93.30&\textbf{93.99}       \\\cline{2-11}
& \multicolumn{1}{|c|}{LPIPS $\downarrow$} & 0.0721 & 0.0977 & 0.1088 & 0.1437 & 0.1491 & 0.1134 & 0.0774 & 0.0694 & \textbf{0.0658}        \\\hline

\multirow{1}{*}{RLE Seg}  & \multicolumn{1}{|c|}{mIoU $\uparrow$} &66.47 & 59.46 & 51.47 & 62.46 & 52.57 & 45.33 & 62.56 & 59.92 & \textbf{68.97}        
\\\hline

\end{tabular}}
\end{table}

\begin{table}[t]
\centering
	\caption{
        Ablation study of the proposed EndoUIC on the EndoUIC dataset. Specifically, we (i) replace the restoration DiT with the original U-Net architecture, (ii)~remove the API block, and (iii) remove the GPS block.
	}
 	\centering
        \resizebox{0.7\textwidth}{!}{
 \label{tab:ablation}
\begin{tabular}{cc|ccccccccc}
\hline
\multicolumn{2}{c|}{Diffusion Trans} & \multicolumn{1}{c}{\XSolidBrush} & \multicolumn{1}{c}{\Checkmark} & \multicolumn{1}{c}{\XSolidBrush} & \multicolumn{1}{c}{\XSolidBrush} & \multicolumn{1}{c}{\Checkmark} & \multicolumn{1}{c}{\Checkmark} & \multicolumn{1}{c}{\XSolidBrush} & \multicolumn{1}{c}{\Checkmark}  \\\hline
\multirow{2}{*}{Prompt} & \multicolumn{1}{|c|}{API} & \multicolumn{1}{c}{\XSolidBrush} & \multicolumn{1}{c}{\XSolidBrush} & \multicolumn{1}{c}{\Checkmark} & \multicolumn{1}{c}{\XSolidBrush} & \multicolumn{1}{c}{\Checkmark} & \multicolumn{1}{c}{\XSolidBrush} & \multicolumn{1}{c}{\Checkmark} & \multicolumn{1}{c}{\Checkmark} \\\cline{2-10}
& \multicolumn{1}{|c|}{GPS} & \multicolumn{1}{c}{\XSolidBrush} & \multicolumn{1}{c}{\XSolidBrush} & \multicolumn{1}{c}{\XSolidBrush} & \multicolumn{1}{c}{\Checkmark} & \multicolumn{1}{c}{\XSolidBrush} & \multicolumn{1}{c}{\Checkmark} & \multicolumn{1}{c}{\Checkmark} & \multicolumn{1}{c}{\Checkmark}  \\ \hline
\multicolumn{2}{c|}{PSNR $\uparrow$} & 28.18 & 29.16 & 28.45 & 28.47 & 29.31 & 29.42 & 28.78 & \textbf{29.65}   \\ 
\multicolumn{2}{c|}{SSIM $\uparrow$} & 95.79 & 95.81 & 96.49 & 96.60 & 94.92 & 95.73 & 96.21 & \textbf{96.80}    \\
\multicolumn{2}{c|}{LPIPS $\downarrow$} & 0.0941 & 0.0735 & 0.858 & 0.0776 & 0.0710 & 0.0682 & 0.0727 & \textbf{0.0655} \\ \hline
\end{tabular}}
\end{table}

\noindent\textbf{LLIE Comparison.} We also compare our method with SOTA LLIE techniques on two publicly available LLIE datasets, with the results presented in Table~\ref{tab:3}. We demonstrate that our method still achieves excellent results even when only one type of illumination degradation occurs. We also conduct a segmentation experiment on red lesions following LLCaps~\cite{bai2023llcaps}, and the superior results further prove the clinical applicability of our EndoUIC.

\noindent\textbf{Ablation Study.} In Table~\ref{tab:ablation}, we remove the API and GPS modules from the prompt architecture, and revert the DiT back to the UNet architecture. In all cases, we observed varying degrees of performance degradation, which further proves the effectiveness of the proposed components.

\noindent\textbf{Feature Clustering.} We also observe the feature clustering of over \& underexposed images in the CEC test set with t-SNE. After passing through the 1st, 2nd, and 3rd prompt blocks, the feature clustering between over \& underexposed images became more distinct and better clustered (Davies-Bouldin Index $1.68 \rightarrow 1.55 \rightarrow 0.53$). After removing the prompt blocks, we find worse clustering results ($2.24 \rightarrow 1.94 \rightarrow 1.87$), indicating that the prompt block can help optimize the feature clustering based on the discriminative illumination information.

\section{Conclusion}
\label{sec:conclusion}
This paper presents EndoUIC, a promptable DiT model for unified illumination correction for WCE. 
The model's ability to navigate through different regions of the parameter space allows for tailored adjustments that address the distinct challenges posed by either overexposed or underexposed images. Furthermore, with the assistance of photographer experts, we customize the CEC dataset tailored for the EC task in WCEs. Extensive experiments conducted on four datasets demonstrate that our EndoUIC surpasses existing SOTA techniques, validating its efficacy in performing endoscopic LLIE and EC tasks. Our proposed approach can be integrated with clinical endoscopy systems, greatly enhancing the visualization, diagnosis, screening, and treatment of GI diseases.

\begin{credits}
\subsubsection{\ackname} This work was supported by Hong Kong RGC GRF 14211420, CRF C4063-18G, NSFC/RGC Joint Research Scheme N\_CUHK420/22; Shenzhen-HK-Macau Technology Research Programme (Type C) STIC Grant 202108233000303; Regional Joint Fund Project 2021B1515120035 (B.02.21.00101) of Guangdong Basic and Applied Research Fund.

\subsubsection{\discintname}
The authors have no competing interests to declare that are relevant to the content of this article.
\end{credits}

\bibliography{Paper-0221}{}
\bibliographystyle{splncs04}

\newpage

\section*{Supplementary Materials for ``EndoUIC: Prompt Diffusion Transformer for  Unified Illumination Correction in Capsule Endoscopy''}

\begin{table}[h]
	\caption{
        Full results of EC comparison with existing methods on our CEC dataset.
	}
 	\centering
	\label{tab:sup1}  
 \resizebox{\textwidth}{!}{	
\begin{tabular}{c|ccccccccc}
\hline
Models & RetinexNet & Zero-DCE & RCTNet & MSEC & FECNet & DRBN-ENC & MIRNetv1 & MIRNetv2 & LA-Net  \\ \hline
PSNR $\uparrow$ & 19.05 & 10.30 & 18.90 & 7.54 & 28.78 & 26.83 & 27.62 & 28.36 & 16.58\\ 
SSIM $\uparrow$ & 76.74 & 65.46 & 65.03 & 38.77 & 92.61 & 90.50 & 93.03 & 93.58 & 66.40 \\ 
LPIPS $\downarrow$ & 0.4522 & 0.5873 & 0.5847 & 0.6561 & 0.1048 & 0.1452 & 0.1252 & 0.1080  & 0.4233 \\ \hline
Models  & LLCaps & PyDiff & PromptIR & LACT & PIP  & CLE & PSENet & DiT & EndoUIC \\ \hline 
PSNR $\uparrow$  & 27.55 & 28.18 & 28.27 & 28.40 & 25.01 & 27.12 & 12.70 & 23.43 & \textbf{29.65}\\ 
SSIM $\uparrow$  & 85.95 & 95.79 & 83.14 & 93.09 & 70.09 & 69.99 & 64.74 & 91.18 & \textbf{96.80}\\ 
LPIPS $\downarrow$ & 0.2366 & 0.0941 & 0.0717 & 0.1103 & 0.1833 & 0.1527 & 0.4973 & 0.1560 & \textbf{0.0655}\\ 
\hline
\end{tabular}}
\end{table}

\begin{table}[h]
	\caption{
        Full results of EC comparison with existing methods on the Endo4IE dataset.
	}
 	\centering
	\label{tab:sup2}  
 \resizebox{\textwidth}{!}{	
\begin{tabular}{c|ccccccccc}
\hline
Models & LMSPEC* & LMSPEC+* & RetinexNet & Zero-DCE & RCTNet & MSEC & FECNet  & DRBN-ENC & MIRNetv2 \\ \hline
PSNR $\uparrow$ &23.97 &23.62 & 16.82 & 14.30 & 23.66 & 22.24 & 24.72  & 7.86 &23.85\\ 
SSIM $\uparrow$ &23.62 &79.97 & 68.78 & 68.15 & 79.59 & 76.05 & 81.84  & 5.14 &82.33 \\ 
LPIPS $\downarrow$ &- &- & 0.3148 & 0.4840 & 0.3041 & 0.2508 & 0.2031 & 0.8516 &0.2376 \\ \hline
Models  & LA-Net & LLCaps & PyDiff & PromptIR & LACT & PIP  & CLE & PSENet & EndoUIC \\ \hline 
PSNR $\uparrow$ & 23.51 & 22.85 & 24.73 & 23.73 & 22.92 & 25.28 & 21.20 & 17.49 & \textbf{25.49}\\ 
SSIM $\uparrow$ & 83.78 & 66.42 & 84.78 & 79.57 & 76.88 & 81.94 & 57.39 & 71.12 & \textbf{85.20}\\ 
LPIPS $\downarrow$ & \textbf{0.1186} & 0.3446 & 0.2148 & 0.2396 & 0.2671 & 0.2150 & 0.3816 & 0.2839 & 0.1937\\ 
\hline
\end{tabular}}
\end{table}

\begin{figure}
\centering
\includegraphics[width=0.9\textwidth, trim=-50 200 50 0]{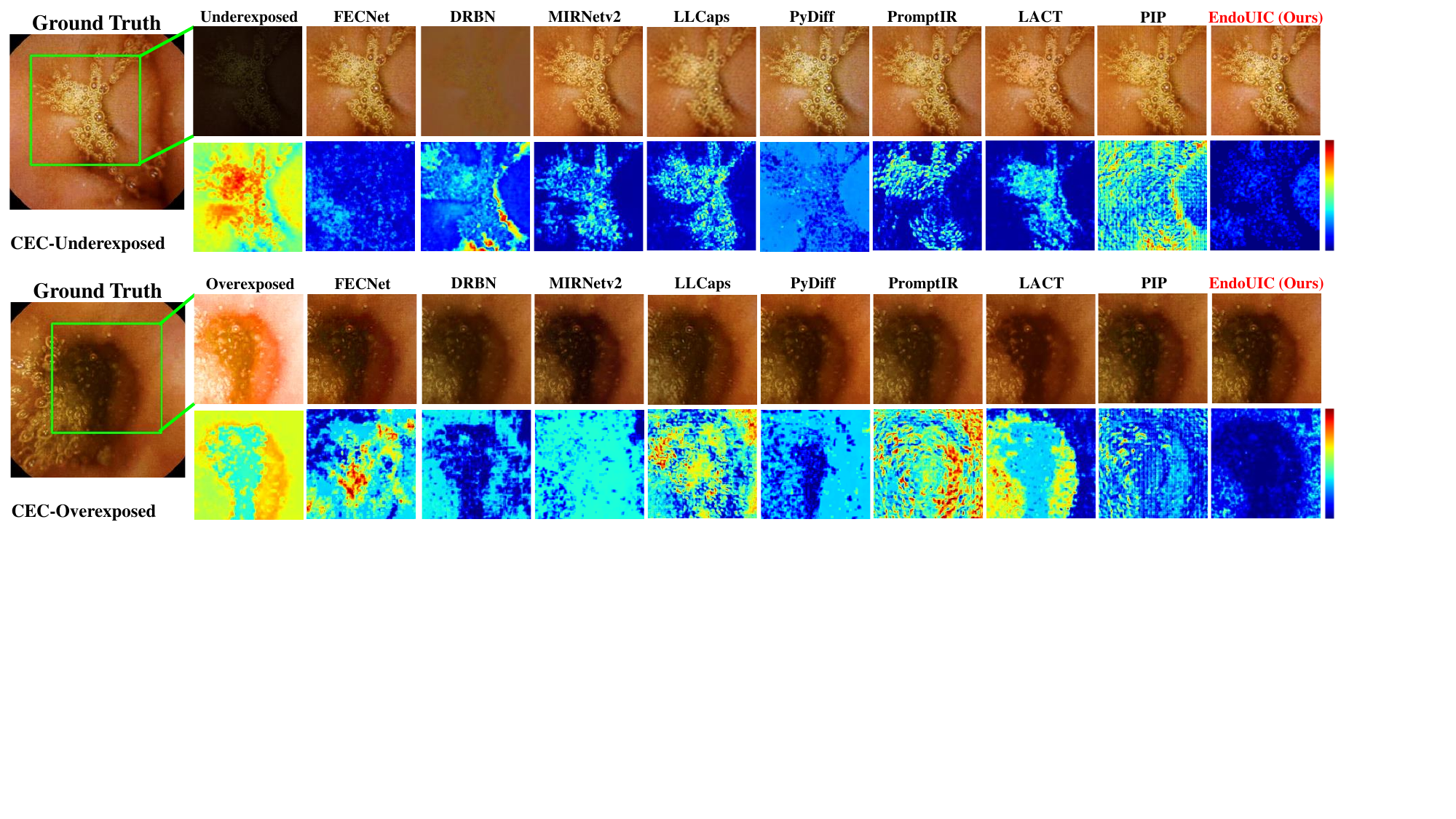}
\caption{The visualization for the EC task on the CEC dataset.} \label{figa1}
\end{figure}

\begin{figure}

\centering
\includegraphics[width=0.8\textwidth, trim=-20 120 0 0]{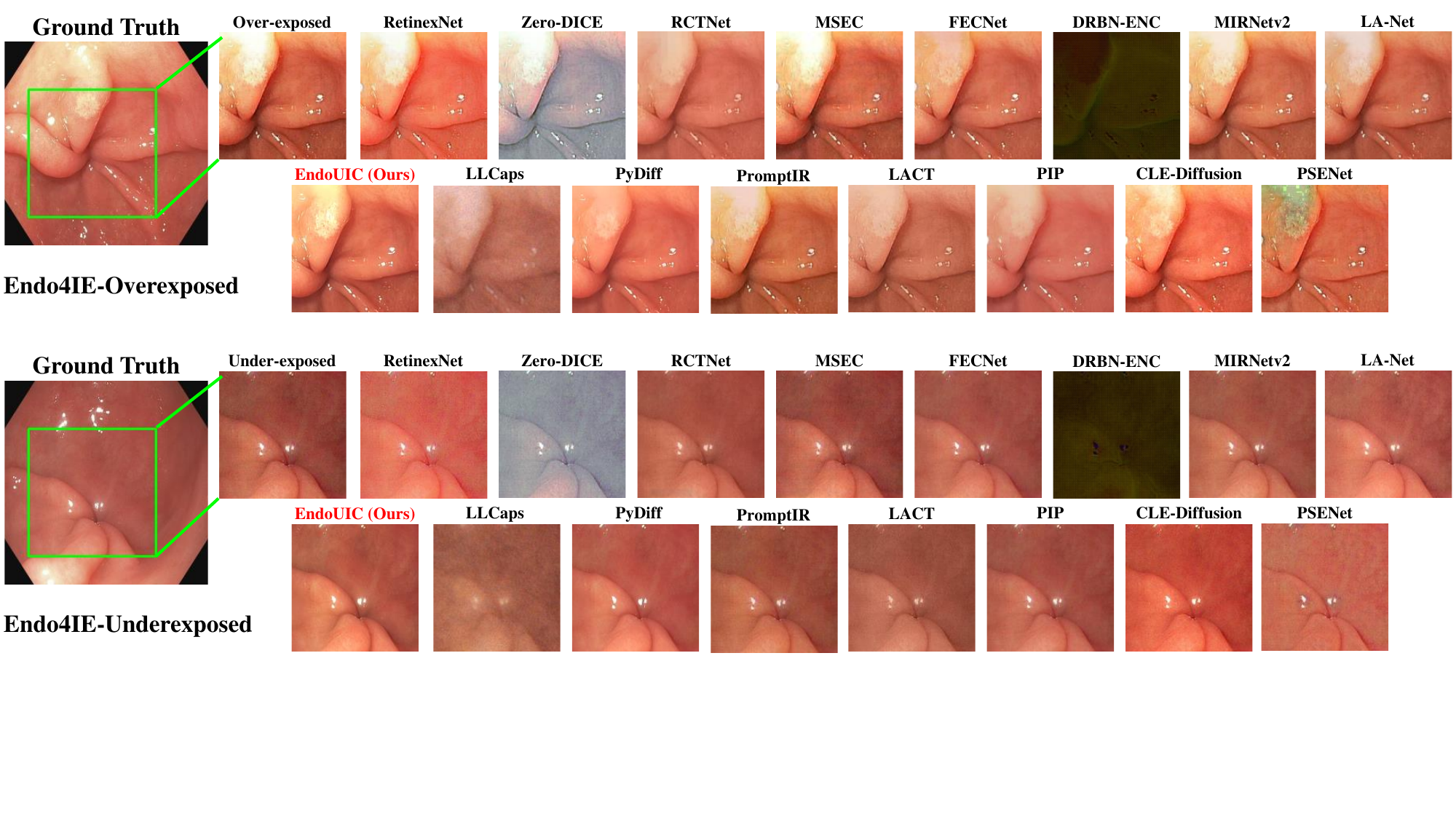}
\caption{The visualization for the EC task on the Endo4IE dataset.} \label{figa2}
\end{figure}

\begin{figure}
\centering
\includegraphics[width=0.75\textwidth, trim=0 600 0 0]{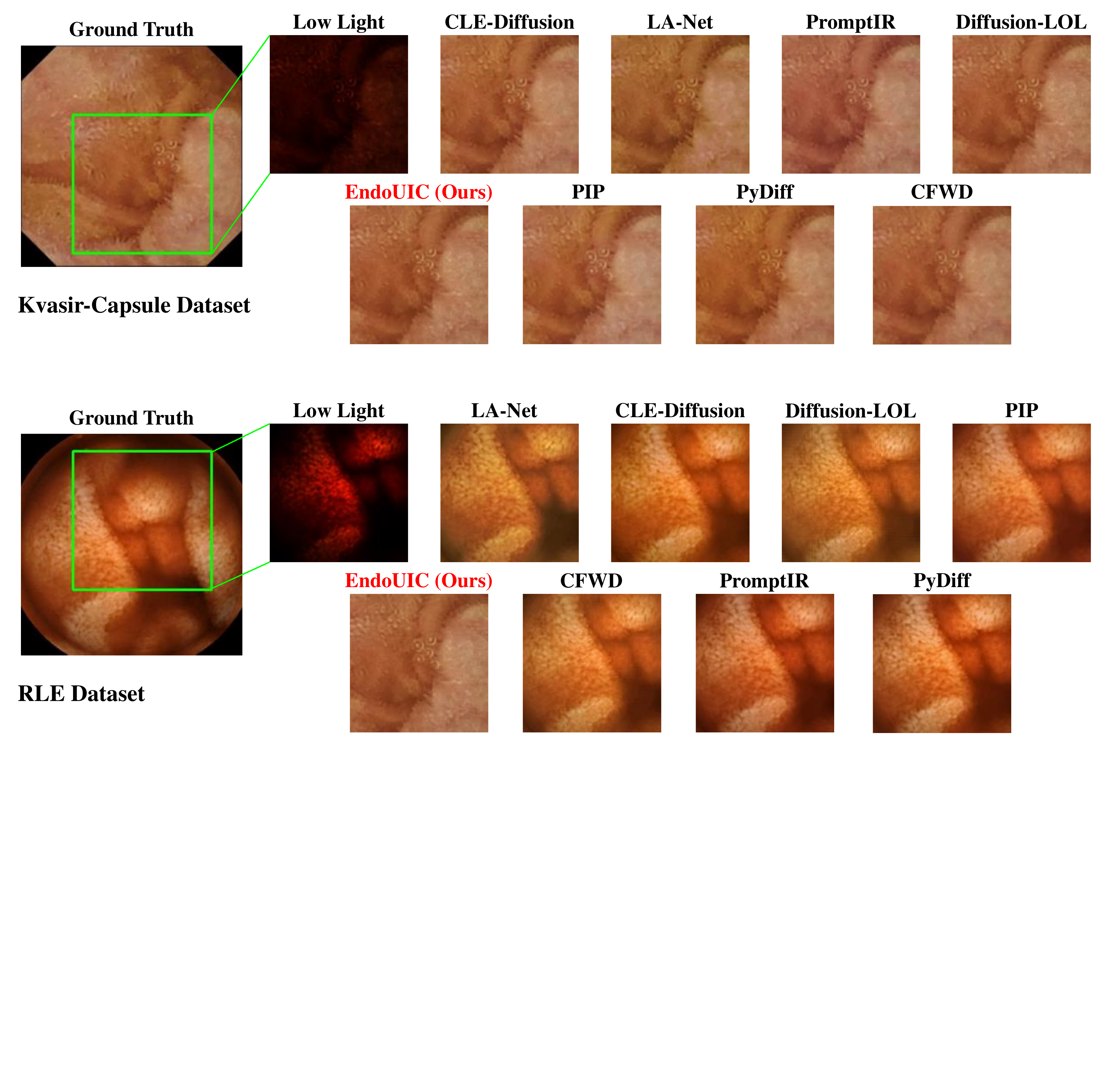}
\caption{The visualization for the LLIE task on Kvasir-Capsule and RLE datasets.} \label{figa3}
\end{figure}

\end{document}